\documentclass[11pt,prd,aps,showpacs,eqsecnum,floatfix,nofootinbib,preprint,tightenlines]{revtex4}
\usepackage{graphicx}				
\usepackage{latexsym}
\usepackage{epsf}
\usepackage{graphicx}
\usepackage{slashed}
\usepackage{amsmath}
\usepackage{amssymb}

\usepackage{color}
\usepackage{colordvi}
\usepackage{multirow}
\newcommand{\be}{\begin{equation}}
\newcommand{\ee}{\end{equation}}
\newcommand{\bea}{\begin{eqnarray}}
\newcommand{\eea}{\end{eqnarray}}
\newcommand{\ba}{\begin{array}}
\newcommand{\ea}{\end{array}}
\newcommand{\nn}{\nonumber}

\newcommand{\pref}[1]{(\ref{#1})}
\newcommand{\np}{N\pi}

\newcommand{\ENq}{E_{N,\vec{q}}}

\newcommand{\Epiq}{E_{\pi,{\vec{q}}}}

\newcommand{\GA}{G_{\rm A}}
\newcommand{\GP}{G_{\rm P}}
\newcommand{\GPt}{\tilde{G}_{\rm P}}

\newcommand{\DG}{\Delta G}

\begin{document}
\renewcommand{\thefootnote}{$*$}

\preprint{HU-EP-19/38}

\title{$N\pi$-states and the projection method for the nucleon axial and pseudoscalar form factors}

\author{Oliver B\"ar$^{a}$} 
\affiliation{$^a$Institut f\"ur Physik,
\\Humboldt Universit\"at zu Berlin,
\\12489 Berlin, Germany\\}

\begin{abstract}
The RQCD collaboration
proposed a projection method to remove the excited state contamination in lattice QCD calculations of nuclear form factors.
The effectiveness of this method in removing the two-particle nucleon-pion-state contamination is examined using chiral perturbation theory. It is shown that the projection method has practically no impact in the calculation of the axial and induced pseudoscalar form factors. In the pseudoscalar form factor the projection method strongly enhances the nucleon-pion-state contamination.
The generalized Goldberger-Treiman relation is satisfied even though large nucleon-pion-state contaminations are present in individual form factors. Therefore, the projection method is not a solution to the excited state problem in nucleon
form factor calculations.
\end{abstract}

\maketitle

\renewcommand{\thefootnote}{\arabic{footnote}} \setcounter{footnote}{0}

\newpage

%
\section{Introduction}
%

Lattice QCD results can have a valuable impact on particle physics areas beyond QCD. 
For example, Ref.\ \cite{Kronfeld:2019nfb}, one of a series of whitepapers from the USQCD collaboration, discusses the opportunities for lattice QCD in neutrino-oscillation physics. Lattice QCD is in principle able to provide various observables, among others the vector and axial vector form factors of the nucleon. Lattice calculations of these form factors have a long history and are straightforward to carry out. However, in order to be phenomenologically relevant the lattice results need to have small and reliable statistical and systematic errors such that the total error is at the percent level. Currently this is still challenging to achieve. 

Even though the lattice techniques for the calculation of nucleon form factors are well-established, recent calculations of the axial form factors display an unexpected and puzzling behavior: The partially conserved axial vector current (PCAC) relation implies the so-called generalized Goldberger-Treiman (gGT) relation between the axial and pseudoscalar form factors, but the lattice results for the form factors strongly violate this relation \cite{Rajan:2017lxk,Tsukamoto:2017fnm,Jang:2018lup,Ishikawa:2018rew}. As pointed out by the RQCD collaboration \cite{Bali:2018qus}, this so-called {\em PCAC puzzle} is surprising since the PCAC relation is fulfilled rather well on the level of the correlation functions. This strongly suggests a large excited state contamination as the source for violating the gGT relation. 

As a remedy RQCD proposed a simple projection method to remove the large excited state contamination from the correlation functions \cite{Bali:2018qus}. 
Numerical tests of this method show that the gGT relation is indeed satisfied if the projection method is applied. Still, the results are not fully conclusive.  A large excited state contamination in the induced pseudoscalar form factor is essentially unaffected by the projection, and the theoretically expected pion pole dominance for this form factor is as badly violated as before applying the projection method.

The excited state contamination due to two-particle nucleon-pion ($N\pi$) states has recently been studied in Refs.\ \cite{Bar:2018xyi,Bar:2019gfx,Bar:2019zkx} within chiral perturbation theory (ChPT). It has been shown that the observed violation of the gGT relation can be explained by a large $N\pi$ contamination in the induced pseudoscalar form factor. The $N\pi$ contamination in the axial and pseudoscalar form factors have a much smaller influence, in particular for small momentum transfer. These findings are apparently in contradiction with the results in \cite{Bali:2018qus}: How can the projection method solve the PCAC puzzle if it has essentially no impact on the $N\pi$ contamination in the induced pseudoscalar form factor?

In this paper we extend the ChPT results in Ref. \cite{Bar:2018xyi,Bar:2019gfx} to the projected axial vector current and pseudoscalar density. The results provide an analytical understanding for the effectiveness of the projection method in removing the $N\pi$ state contamination and explain the apparent contradiction mentioned before.

The main results of this paper are easily summarized. The projection method removes only part of the $N\pi$ contamination in the pseudoscalar form factor. The remaining part is larger than the original one because the subtraction upsets a delicate cancellation in the $N\pi$ contamination. As a consequence the pseudoscalar form factor obtained with the projection method is largely overestimated by the remaining $N\pi$ contamination. This overestimation compensates the underestimation of the induced pseudoscalar form factor and the gGT relation is indeed satisfied. 
Obviously this is not the desired solution to the PCAC puzzle:  Applying the projection method we are left with two form factors instead of one afflicted with a large excited state contamination. Comparing the ChPT results with the lattice data of \cite{Bali:2018qus} strongly supports this conclusion.

This paper relies heavily on the results in Ref.\ \cite{Bar:2018xyi,Bar:2019gfx}, and the reader is assumed to be familiar with these references. The general ideas behind ChPT calculations of the  $\np$-state contamination in nucleon observables have been recently reviewed in \cite{Bar:2017kxh,Bar:2017gqh} and are not repeated here. 
%
\section{Nucleon axial form factors}
%
\subsection{Basic definitions}
We are interested in the matrix elements of the local iso-vector axial vector current $A^a_{\mu}(x)$ and pseudoscalar density $P^a(x)$  between single-nucleon (SN) states $| N(p,s)\rangle$ of definite momenta and spin,
\begin{eqnarray}
\langle N(p',s')|A_{\mu}^a(0)|N(p,s)\rangle & = & \bar{u}(p',s')\left(\gamma_{\mu}\gamma_5 \GA(Q^2) - i \gamma_5\frac{Q_{\mu}}{2M_N}\GPt(Q^2)\right)\frac{\sigma^a}{2}u(p,s)\,,\label{MElements1}\\
m_q \langle N(p',s')|P^a(0)|N(p,s)\rangle & = & m_q \GP(Q^2)\bar{u}(p',s')\gamma_5 \frac{\sigma^a}{2}u(p,s)\,.\label{MElements2}
\end{eqnarray}
The right hand side shows the form factor decomposition of the matrix elements. 
$m_q$ denotes the mass of the up and down quarks which we assume to be degenerate.
$u(p,s)$ is an isodoublet Dirac spinor with momentum $p$ and spin $s$, and in euclidean space-time the four-momentum transfer $Q_{\mu}$ is given by
\begin{equation}
Q_{\mu}=(i E_{N,\vec{p}^{\,\prime}} - i E_{N,\vec{p}},\vec{q})\,,\qquad \vec{q} = \vec{p}^{\,\prime}-\vec{p}\,.
\end{equation}
In euclidean  (lattice) QCD the form factors are computed for space-like momentum transfers $Q^2>0$, with 
$Q^2=(\vec{p}^{\,\prime}-\vec{p})^2 - (E_{N,\vec{p}^{\,\prime}} - E_{N,\vec{p}})^2$ and and the nucleon energy $E_{N,\vec{p}}^{\,2}=\vec{p}^{\,2}+M_N^2$.

The two matrix elements are decomposed into three form factors: the axial form factor $\GA(Q^2)$, the induced pseudoscalar form factor $\GPt(Q^2)$, and the pseudoscalar form factor $\GP(Q^2)$. These three form factors are not independent. Taking the PCAC relation
\begin{equation}\label{PCACrelation}
\partial_{\mu}A^a_{\mu}(x) = 2 m_q P^a(x)
\end{equation}
between SN states leads to the gGT relation
\begin{equation}\label{pcacformfactorlevel}
2M_N G_{\rm A}(Q^2) -\frac{Q^2}{2M_N} \GPt(Q^2) = 2 m_q \GP(Q^2)
\end{equation}
for the three form factors.\footnote{Ref.\ \cite{Bali:2018qus} refers to it as the PCAC$_{\rm FF}$ relation.} 

Considering \pref{pcacformfactorlevel} in the limit of vanishing momentum transfer and pion mass one can conclude that both  $\GPt(Q^2)$ and $ \GP(Q^2)$ are dominated by a pion pole for small $Q^2$. For $Q^2$ close to $-M_{\pi}^2$
one can derive the expressions\footnote{See appendix B of Ref.\ \cite{Sasaki:2007gw} for a quick derivation.} 
\begin{eqnarray}
\GPt^{\rm ppd}(Q^2)  &=& \frac{4M_N^2}{Q^2+M_{\pi}^2} G_{\rm A}(Q^2)\,,\label{ppd1}\\
2 m_q \GP^{\rm ppd}(Q^2) &=& \frac{2M_NM_{\pi}^2}{Q^2+M_{\pi}^2} G_{\rm A}(Q^2)\,,\label{ppd2}
\end{eqnarray}
for the form factors, which are called the PPD model results.

The standard procedure to compute the form factors in lattice QCD is based on evaluating various 2- and 3-point (pt) functions. The nucleon 2-pt function is given by
\begin{equation}\label{Def2ptfunc}
C_2(\vec{p},t)= \int d^3x \,e^{i\vec{p}\vec{x}}\, \Gamma_{\beta\alpha}\langle N_{\alpha}(\vec{x},t)\overline{N}_{\beta}(0,0)\rangle \,.
\end{equation}
$N,\overline{N}$ denote interpolating fields of the nucleon. We assume them to be given by the standard 3-quark operators \cite{Ioffe:1981kw,Espriu:1983hu} (either point like or smeared) that have been mapped to ChPT \cite{Wein:2011ix,Bar:2015zwa,Bar:2013ora}. The projector $\Gamma=(1+\gamma_4)(1+i \gamma_5\gamma_3)/4$ acts on spinor space and projects onto the positive-parity sector.

The nucleon 3-pt functions are typically computed with the nucleon at the sink being at rest, $\vec{p}^{\,\prime}=0$, and the 
third isospin component is chosen as $a=3$. Thus, the nucleon 3-pt functions we consider are given by
\begin{eqnarray}
C_{3,A^3_{\mu}}(\vec{q},t,t')&=&\int d^3x\int d^3y \,e^{i\vec{q}\vec{y}}\, \Gamma_{\beta\alpha}\langle N_{\alpha}(\vec{x},t) A_{\mu}^3(\vec{y},t')\overline{N}_{\beta}(0,0)\rangle\,,\label{C3ptAmu}\\
C_{3,P^3}(\vec{q},t,t')&=&\int d^3x\int d^3y \,e^{i\vec{q}\vec{y}}\, \Gamma_{\beta\alpha}\langle N_{\alpha}(\vec{x},t) P^3(\vec{y},t')\overline{N}_{\beta}(0,0)\rangle\,.\label{C3ptP}
\end{eqnarray}
The euclidean times $t$ and $t'$ denote the source-sink separation and the operator insertion time, respectively.
With the 2-pt and 3-pt functions the generalized ratios ($\mu=1,\ldots 4,P$)
\begin{eqnarray}\label{DefRatio}
R_{\mu}(\vec{q},t,t')& =&\frac{C_{3,X^3_{\mu}}(\vec{q},t,t')}{C_2(0,t)}\sqrt{\frac{C_2(\vec{q},t-t')}{C_2(0,t-t')}\frac{C_2(\vec{0},t)}{C_2(\vec{q},t)}\frac{C_2(\vec{0},t')}{C_2(\vec{q},t')}}\,,
\end{eqnarray}
are defined.  As a short hand notation $\mu=P$ refers to the ratio with the pseudoscalar 3-pt function \pref{C3ptP}. 
The ratios are defined in such a way that, in the asymptotic limit $t,t', t-t'\rightarrow \infty$, they converge to constant asymptotic values, 
\begin{equation}\label{AsympValues}
R_{\mu}(\vec{q},t,t') \longrightarrow \Pi_{{\mu}}(\vec{q})\,.
\end{equation}
The form factors are obtained from these constant values. For example, the pseudoscalar form factor $\GP(Q^2)$ is directly proportional to $\Pi_{\rm P}(\vec{q})$,
\begin{equation}
\Pi_{\rm P}(\vec{q}) =  \frac{q_3}{\sqrt{2E_{N,\vec{q}}(M_N+ E_{N,\vec{q}})}}\,\GP(Q^2)\,.\label{AsympValueP}
\end{equation}
The proportionality factor is a simple kinematical factor that is easily computed and removed from $\Pi_{\rm P}(\vec{q})$.
The axial form factors $\GA(Q^2)$ and $\GPt(Q^2)$ can be computed analogously, although in general one has to solve a linear system to extract the two form factors from two independent asymptotic values $\Pi_{{\mu}}(\vec{q})$.\footnote{The expressions for $\Pi_{{\mu}}(\vec{q})$ in terms of the axial form factors are given in eqs.\ (2.16), (2.17) of Ref.\ \cite{Bar:2019gfx}, for example.}

Although this method works in principle, in practice one only has access to the ratios $R_{\mu}(\vec{q},t,t')$ at time separations $t,t'$ that are far from being asymptotically large. In that case the correlation functions and the ratios not only contain the SN ground-state contribution, but also contributions of excited states. This excited-state contamination enters the calculation of the form factors: Instead of the true form factors we obtain {\em effective} form factors including an excited-state contamination. These effective form factors are expected to be of the form\footnote{For brevity we introduce the notation $G_{\tilde{\rm P}}=\tilde{G}_{\rm P}$.} 
\begin{eqnarray}
G^{\rm eff}_{\rm X}(Q^2,t,t')\, = \,G_{\rm X}(Q^2)\bigg[ 1 + \DG_{\rm X}(Q^2,t,t')\bigg],\quad X\,=\,A,P,\tilde{P}\,.\label{EffFF}
\end{eqnarray}
The excited-state contribution $\DG_{\rm X}(Q^2,t,t')$ vanishes for $t,t',t-t'\rightarrow \infty$. 

The effective form factors depend on both the source-sink separation $t$ and the operator insertion time $t'$. As an estimator for the form factor we can introduce the plateau estimate $G_{\rm X}^{\rm plat}(Q^2,t)$ that, for a given source-sink separation $t$, fixes $t'$ to the value that minimizes the deviation from the true form factor. Alternatively one can define the midpoint estimate $G_{\rm X}^{\rm mid}(Q^2,t)=G_{\rm X}^{\rm eff}(Q^2,t,t'=t/2)$. Both are functions of the momentum transfer and $t$. In practice the difference between the two estimators is small, at least for small momentum transfers. 

Instead of the standard current and density the projection method proposed in \cite{Bali:2018qus} uses
\begin{eqnarray}
A_{\mu}^{a,\perp}(x) &=& A^a_{\mu}(x) - \frac{\overline{p}_{\mu}\overline{p}_{\nu}}{\overline{p}^2}A^a_{\nu}(x)\,,\label{DefAperp}\\
P^{a,\perp} (x)& =& P^a(x) - \frac{1}{2m_q}  \frac{\overline{p}_{\mu}\overline{p}_{\nu}}{\overline{p}^2} \partial_{\mu}A^a_{\nu}(x)\,,\label{DefPperp}
\end{eqnarray}
with $\overline{p}_{\mu}=(p^{\prime}_{\mu}+p_{\mu})/2$ being the mean of the initial and final nucleon momenta. By construction the projected current and density satisfy the PCAC relation \pref{PCACrelation} and the contraction of $\overline{p}_{\mu}$ with the SN matrix element in \pref{MElements1} vanishes \cite{Bali:2018qus}. Therefore, the ratios $R^{\perp}_{\mu}(\vec{q},t,t')$, formed with the 3-pt functions of the projected current and density, have the same constant asymptotic values as the original ratios in \pref{AsympValues}. However, the effective form factors $G^{\rm eff}_{\rm X^{\perp}}(Q^2,t,t')$ obtained from $R^{\perp}_{\mu}(\vec{q},t,t')$ at finite $t,t'$ differ because the excited-state contaminations are in general different,
\begin{equation}
\DG_{\rm X^{\perp}}(Q^2,t,t')\neq\DG_{\rm X}(Q^2,t,t')\,.
\end{equation}
Therefore, the plateau and midpoint estimators also differ depending on which currents or densities are used.

As a quantitative measure for violations of the gGT relation one can introduce the dimensionless ratio \cite{Rajan:2017lxk,Bali:2018qus}  
\begin{equation}\label{DefrPCAC}
r^{\rm est}_{\rm PCAC}(Q^2,t) = \frac{Q^2}{4 M_N^2} \frac{\GPt^{\rm est}(Q^2,t)}{\GA^{\rm est}(Q^2,t)} + \frac{m_q}{ M_N} \frac{\GP^{\rm est}(Q^2,t)}{\GA^{\rm est}(Q^2,t)}\,,
\end{equation}
for both the plateau and midpoint estimator, and, analogously, $r^{\perp,{\rm est}}_{\rm PCAC}(Q^2,t)$. In the limit $t\longrightarrow \infty$ these ratios assume the constant value $1$. This is nothing but the gGT relation \pref{pcacformfactorlevel}. For finite $t$ the excited-state contamination in the form factor estimators result in deviations from 1. One typically finds values smaller than 1, and the deviation increases the smaller the momentum transfer is \cite{Rajan:2017lxk,Bali:2018qus}.
%
\section{Nucleon-pion excited states}
%
\subsection{Preliminaries}
In lattice simulations with pion masses as small as in Nature one can expect two-particle $N\pi$ states to cause the dominant excited-state contamination, 
\begin{equation}\label{appDeltaG}
\DG_{\rm X}(Q^2,t,t') \approx \DG^{N\pi}_{\rm X}(Q^2,t,t')\,,
\end{equation}
provided the time separations $t,t'$ are sufficiently large. This expectation rests on the naive observation that the energy gaps between the $N\pi$ states and the SN ground state are smaller than those from resonance states and other (heavier) multi-hadron states. For this to happen not only the pion mass needs to be small, sufficiently large spatial volumes are also necessary such that the discrete spatial momenta imply sufficiently small energies for the lowest-lying $N\pi$ states. Volumes with $M_{\pi}L\simeq 4$, typically used in lattice simulations, fulfill this criterion \cite{Bar:2017kxh}.

Provided we are in the regime where \pref{appDeltaG} holds we can use ChPT to get an estimate for the $N\pi$-state contamination  $\DG^{N\pi}_{\rm X}$ for all three form factors. For the axial form factors the calculation to LO is given in \cite{Bar:2018xyi}, the analogous result for the pseudoscalar form factor can be found in \cite{Bar:2019gfx}. With these results it is straightforward to derive the corresponding results for the projected current and density, i.e.\ the contaminations $\DG^{N\pi}_{\rm X^{\perp}}$.

\subsection{$N\pi$-state contribution to nucleon 3-pt functions}\label{ssect:Npicontr}

Performing the standard spectral decomposition in $C_{3,{\mu}}(\vec{q},t,t')$ defined in eq.\ \pref{C3ptAmu}, \pref{C3ptP}, the 3-pt function is found to be a sum of various contributions,
\begin{eqnarray}
C_{3,\mu}(\vec{q},t,t') & = & C^N_{3,\mu}(\vec{q},t,t') + C^{\np}_{3,\mu}(\vec{q},t,t')+\ldots\,. \label{C3muspecDecomp}
\end{eqnarray}
The first two terms on the right hand side refer to the SN and the $N\pi$ contributions. The ellipsis refers to omitted contributions which we assume to be small in the following. Provided the SN contribution is nonzero we may write \pref{C3muspecDecomp} as 
\begin{eqnarray}
C_{3,\mu}(\vec{q},t,t')& =& C^N_{3,\mu}(\vec{q},t,t')\bigg(1+ Z_{\mu}(\vec{q},t,t')\bigg)\label{DefC3Npcontr}\,.
\end{eqnarray}
Thus, $Z_{\mu}$ denotes the ratio $C^{\np}_{3,\mu}(\vec{q},t,t')/C^N_{3,\mu}(\vec{q},t,t')$. With our kinematical setup $\vec{p}^{\,\prime}=0$ the generic form for $Z_{\mu}(\vec{q},t,t')$ is found as \cite{Bar:2018xyi},
\begin{eqnarray}
Z_{\mu}(\vec{q},t,t') & = & a_{\mu}(\vec{q}) e^{-\Delta E(0,-\vec{q}) (t-t')}+ \tilde{a}_{\mu}(\vec{q}) e^{-\Delta E(\vec{q},-\vec{q})t'} \nn \\[2ex]
& & + \sum_{\vec{k}} b_{\mu}(\vec{q},\vec{k}) e^{-\Delta E(0,\vec{k}) (t-t')}+\sum_{\vec{k}}\tilde{b}_{\mu}(\vec{q},\vec{k}) e^{-\Delta E(\vec{q},\vec{k}) t'} \nn\\
& & +  \sum_{\vec{k}} c_{\mu}(\vec{q},\vec{k}) e^{-\Delta E(0,\vec{k}) (t-t')}e^{-\Delta E(\vec{q},\vec{k}) t'}\,.\label{DefZmu}
\end{eqnarray}
The sum runs over all pion momenta $\vec{k}$ that are compatible with the boundary conditions imposed for the spatial volume. The nucleon momentum $\vec{r}$ is fixed to $\vec{r}=-\vec{q}-\vec{k}$ by momentum conservation. To LO in ChPT the energy gaps $\Delta E(\vec{q},\vec{k})$ between the SN ground-state and the $N\pi$ states are obtained by ignoring the (small) nucleon-pion interaction energies, i.e.\ 
\begin{equation}\label{Egap2pt}
\Delta E(\vec{q},\vec{k}) = E_{\pi,\vec{k}} + E_{N,\vec{q}+\vec{k}} - \ENq\,.
\end{equation}
The coefficients $a_{\mu}(\vec{q}), \tilde{a}_{\mu}(\vec{q}),b_{\mu}(\vec{q},\vec{k}) ,\tilde{b}_{\mu}(\vec{q},\vec{k}), c_{\mu}(\vec{q},\vec{k})$ in \pref{DefZmu} are ratios of matrix elements involving the nucleon interpolating fields and either the axial vector current or pseudoscalar density. To obtain ChPT estimates for these coefficients both $C^N_{3,\mu}(\vec{q},t,t')$ and $C^{\np}_{3,\mu}(\vec{q},t,t')$ as well as the ratio needs to be computed in ChPT. To LO this involves twelve 1-loop and three tree diagrams for $C^{\np}_{3,\mu}(\vec{q},t,t')$. For the axial vector 3-pt function this has been done in \cite{Bar:2018xyi}, and the explicit results for the coefficients are given in section IV of this reference.

In order to compute $Z_{P}(\vec{q},t,t')$, the $N\pi$ contribution in the pseudoscalar 3-pt function, one may calculate the same diagrams with the axial vector current replaced by the pseudoscalar density. Alternatively, the result can be obtained by making use of the PCAC relation which relates the pseudoscalar coefficients to the axial vector ones. This route has been followed in Ref.\ \cite{Bar:2019gfx}

With the ChPT results for $Z_{\mu}(\vec{q},t,t')$ at hand it is straightforward to compute the analogous $N\pi$ contributions associated with the projected current and density, $Z^{\perp}_{\mu}(\vec{q},t,t')$. It is defined with \pref{DefAperp}, \pref{DefPperp} in the 3-pt functions on the left hand side in eq.\ \pref{DefC3Npcontr}. $Z^{\perp}_{\mu}(\vec{q},t,t')$ has the same form as $Z_{\mu}(\vec{q},t,t')$ in \pref{DefZmu} but with coefficients carrying a superscript: $a^{\perp}_{\mu}(\vec{q}), b^{\perp}_{\mu}(\vec{q},\vec{k}) $ etc. $A_{\mu}^{a,\perp}(x) $ is a linear combination of all four $A_{\mu}^{a}(x)$, thus the same holds for the $N\pi$ contributions,
\begin{eqnarray}
Z_{\mu}^{\perp}(\vec{q},t,t') &=& Z_{\mu}(\vec{q},t,t') - \Delta Z_{\mu}(\vec{q},t,t')\,,\label{DefZmuperp}\\
\Delta Z_{\mu}(\vec{q},t,t')&=& \sum_{\nu}\frac{\overline{p}_{\mu}\overline{p}_{\nu}}{\overline{p}^2}\,  r^{\nu}_{\mu}\!(\vec{q})\, Z_{\nu}(\vec{q},t,t')\,.\label{DefDeltaZmu}
\end{eqnarray}
The newly introduced $ r^{\nu}_{\mu}\!(\vec{q})$ denotes the ratio of SN contributions in the 3-pt function,
\begin{equation}
r^{\nu}_{\mu}\!(\vec{q}) = \frac{C^N_{3,\nu}(\vec{q},t,t')}{C^N_{3,\mu}(\vec{q},t,t')}\,,
\end{equation}
and to LO it is readily obtained with the results in eqs.\ (4.2), (4.3) of Ref.\ \cite{Bar:2018xyi}. Note that the time dependence cancels in the ratio on the right hand side, so $r^{\nu}_{\mu}$ is a function of the momentum $\vec{q}$ only. 

The calculations in \cite{Bar:2018xyi} are performed in the  covariant formulation of Baryon ChPT \cite{Gasser:1987rb,Becher:1999he}. The results for the coefficients are rational functions involving the energies and masses of the nucleon and pion, and the expressions are 
fairly cumbersome in full covariant form. However, they simplify significantly if we perform the non-relativistic (NR) expansion of the nucleon energy,
\begin{equation}\label{NRexpansion}
\ENq = M_N+\frac{\vec{q}^{\,2}}{2M_N} +\ldots\, \,,
\end{equation}
and keep the first two terms only. For practical uses this approximation is expected to be sufficient. For example, the NR expansion for the coefficients $a_{k}(\vec{q})$ with spatial index $\mu= k=1,2,3,$ reads
\begin{equation}\label{NRexpak}
a_{k}(\vec{q}) = a^{\infty}_{k}(\vec{q}) +\frac{E_{\pi,\vec{q}}}{M_N} a^{\rm corr}_{k}(\vec{q})\,,
\end{equation}
and the results for $a^{\infty}_{k}(\vec{q}),a^{\rm corr}_{k}(\vec{q})$ are given in \cite{Bar:2018xyi}, eqs.\ (4.14) and (4.16). Analogous expressions hold for the other coefficients.

The NR expansion is slightly different for the coefficients with $\mu = 4$. The reason is that the SN contribution in the 3-pt function has a different non-relativistic limit for $\mu=k$ and $\mu=4$. For the latter one finds \cite{Bar:2018xyi}
\begin{eqnarray}
C^N_{3,\mu=4}(\vec{q},t,t') & = &\left[g_A \frac{M_{\pi}^2 q_3}{2 \Epiq^2 M_N}+{\rm O}\left(\frac{1}{M_N^3}\right)\right] e^{-M_N(t-t')}e^{-E_{N,\vec{q}}t'}\,.\label{CN34}
\end{eqnarray}
Thus, the leading term is O$(q_3/M_N)$ suppressed. On the other hand, the expansion of the SN contribution $C^N_{3,k}$  and the $N\pi$ contributions $C^{N\pi}_{3,\mu}$ both start with O(1). The coefficients  in $Z_{\mu}$ are defined by the ratio $C^{N\pi}_{3,\mu}/C^N_{3,\mu}$, thus, for $\mu=4$, the  inverse power $1/M_N$ in the SN contribution shifts the NR expansion of the ratio such that powers linear in the nucleon mass appear. Thus, in contrast to \pref{NRexpak} it is more appropriate to define \cite{Bar:2018xyi}
\begin{equation}\label{NRexpa4}
a_4(\vec{q}) = \frac{M_N}{E_{\pi,q}} a^{\infty}_{4}(\vec{q}) + a^{\rm corr}_{4}(\vec{q})\,.
\end{equation}

In the following we label the $N\pi$ contribution with $Z^{\infty}_{\mu}, Z^{\perp,\infty}_{\mu}$ if the leading NR results are used for the coefficients entering it. It turns out that this leading contribution suffices to qualitatively understand the impact of the projection method, so it is useful to quote these results explicitly.

To obtain the NR limit results $\Delta Z^{\infty}_{\mu}$ it is sufficient to expand $r^{\nu}_{\mu}\!(\vec{q})$ and $\overline{p}_{\mu}\overline{p}_{\nu}/\overline{p}^2$ and consistently drop higher order terms. For the ratio $r^{\nu}_{\mu}\!(\vec{q})$ we use the results in eqs.\ (4.2), (4.3) in \cite{Bar:2018xyi} and obtain ($k,l,=1,2,3$)
\be
r^{\nu}_{\mu}\!(\vec{q})=\left\{
\ba{lcl}
{\rm O}(1) & &
 \mu=\nu=4\,\, {\rm and} \,\, 
  \mu=k, \nu=l\,,\\
{\rm O}(q_k/M_N) & \,{\rm for }\, & \mu=k, \nu=4\,,\\
{\rm O}(M_N/q_k) &  & \mu=4, \nu=k\,.
\ea
\right.
\ee
Recalling the definition $\overline{p}_{\mu}=(p^{\prime}_{\mu}+p_{\mu})/2$ we also find
\be\label{NRExpProjector}
\frac{\overline{p}_{\mu}\overline{p}_{\nu}}{\overline{p}^2} =\left\{
\ba{lcl}
{\rm O}(1) & & \mu=\nu=4\,,\\
{\rm O}(q_k/M_N) & {\rm for } & \mu=k, \nu=4\,,\\
{\rm O}(q_kq_l/M_N^2) &  & \mu=k, \nu=l\,.
\ea
\right.
\ee
Putting all this together in \pref{DefDeltaZmu} yields 
\begin{eqnarray}
\Delta Z^\infty_{k}(\vec{q},t,t')&=& 0\,,\\
\Delta Z^\infty_{4}(\vec{q},t,t')&=& Z^\infty_{4}(\vec{q},t,t')\,.\label{DeltaZ4infty}
\end{eqnarray}
Thus, to LO in the NR expansion the $N\pi$ contamination in the 3-pt function with a spatial component of the axial vector is the same for both the projected and the original axial vector current. For the $\mu=4$ component, on the other hand, the projection removes completely  the LO $N\pi$ contamination, $Z^{\perp,\infty}_{4} =0$. These results will be slightly modified if we take into account the next order in the NR expansion. Still, we can expect the projection method to remove the dominant part of the $N\pi$-state contamination in the $A_4^a$ correlation function, and being essentially ineffective for the spatial components $A_k^a$. Qualitatively this pattern has been observed in Ref.\ \cite{Bali:2018qus}.

The $N\pi$ contamination $Z_{P}$ has been worked out in Ref.\ \cite{Bar:2019gfx}. To LO in the NR expansion one finds\footnote{See eq.\ (3.15) in \cite{Bar:2019gfx}.}
\begin{equation}\label{DefZPinf}
Z^{\infty}_{P}(\vec{q},t,t') = Z^{\prime, \infty}_{4}(\vec{q},t,t') + \sum_{k=1}^3\alpha_k Z^{\infty}_{k}(\vec{q},t,t')\,.
\end{equation}
The $\alpha_k$ are a short-hand notation for simple ratios of the spatial momenta components $q_k$ and the pion mass, see eq.\ (3.30) in \cite{Bar:2019gfx} and eq.\ \pref{resalphak} in appendix \ref{AppA}. $Z^{\prime, \infty}_{4}$ denotes the $N\pi$ contamination of the time derivative of the 3-pt function $C^{N\pi}_{3,4}(\vec{q},t,t')$ with respect to the operator insertion time $t'$. The explicit form of this contribution can also be found in \cite{Bar:2019gfx}, section III.C. See also eqs.\ \pref{DefC3Npcontrprime} - \pref{cpP0} in appendix \ref{AppA}.

The computation of $Z^{\perp}_{P}(\vec{q},t,t')$ is analogous to the one of $Z^{\perp}_{\mu}(\vec{q},t,t')$ and given in appendix \ref{AppA}. Here we simply quote the LO result if the NR expansion is performed,
\begin{eqnarray}
\Delta Z^\infty_{P}(\vec{q},t,t')&=& Z^{\prime,\infty}_{4}(\vec{q},t,t')\,,\label{DeltaZPinfty}\\
Z^{\perp,\infty}_{P}(\vec{q},t,t')&=& \sum_{k=1}^3\alpha_k Z^{\infty}_{k}(\vec{q},t,t')\label{ZPinftyperp}\,.
\end{eqnarray}
In the pseudoscalar case the projection method removes only the $Z^{\prime,\infty}_{4}$ part in the $N\pi$ contamination $Z^{\infty}_{P}$ given in \pref{DefZPinf}. The consequences of this partial subtraction are discussed in the next section.

\section{Comparison with RQCD lattice data}

\subsection{Preliminaries}\label{SSect:Prelim}

To LO ChPT the coefficients in $Z_{\mu},Z^{\perp}_{\mu}$ depend on 5 parameters only, and these are known or easily obtained.
Three of these parameters are the extent $L$ of the spatial volume and the nucleon and pion masses. We set $M_{\pi}= 150$ MeV and $M_{\pi}L=3.5$, the measured values for Ensemble VIII  analysed in \cite{Bali:2018qus}.\footnote{See table 1 in that reference for details of the ensemble.} The nucleon mass is fixed by the measured value $M_{\pi}/M_N=0.160$ \cite{Bali:2014nma}. Errors for these values are at the 1\% level and will be ignored since they are to small to play a role in the following.

In addition, two LO low-energy constants (LECs) need to be specified, the chiral limit values of the pseudoscalar decay constant and axial charge. To LO it is consistent to use the phenomenologically known values and we set $f_{\pi}=93$ MeV and $g_A=1.27$.\footnote{For $g_A$ we could also use the measured value  $g_A\approx1.18$ for ensemble VIII \cite{Bali:2014nma}, but the difference is irrelevant for the results of this paper.}
Note that we do not need values for the LECs associated with the nucleon interpolating fields. To LO these drop out in the ratios $R_{\mu},R^{\perp}_{\mu}$.

We also need to specify an upper bound on the pion momentum in the $N\pi$ state to truncate the sum in \pref{DefZmu}. We follow our earlier studies \cite{Bar:2016uoj,Bar:2016jof,Bar:2018xyi,Bar:2019gfx,Bar:2018wco} and choose $|\vec{k}_n|\leq k_{\rm max}$ with $k_{\rm max}/\Lambda_{\chi}= 0.45$, where the chiral scale $\Lambda_{\chi}$ is equal to $4\pi f_{\pi}$. $N\pi$ states with pions satisfying this bound are called low-momentum $N\pi$ states. For these we expect the LO ChPT results to work reasonably well. States with larger pion momenta  are called high-momentum $N\pi$ states. These too contribute to the excited-state contamination. However, choosing all euclidean time separations sufficiently large the contribution of the high-momentum $N\pi$ states is small, and dropping it leads to a truncation error that can be ignored. The results in Refs.\ \cite{Bar:2016uoj,Bar:2016jof} suggest that separations of about 1 fm or larger between the operator and either source or sink are necessary. This corresponds to source-sink separations of 2 fm or larger in the 3-pt functions. 
In some cases, however, significantly smaller source-sink separations are accessible, for instance in the $A^a_4$ correlation function as well as those that enter the calculation of the induced pseudoscalar form factor $\GPt$ \cite{Bar:2018xyi}.

The lattice data of \cite{Bali:2018qus} we compare to in the following were generated for a source-sink separation $t=1.07$ fm.\footnote{I thank the RQCD collaboration, in particular T.\ Wurm, for sending me the data. Data for two smaller sink separations $t\approx 0.85$ fm and $t\approx 0.64$ fm are also available but seem too small for the ChPT analysis in this paper.} At such small $t$ we would not be surprised if the high-momentum $N\pi$ states were not sufficiently suppressed for the truncation error to be negligible. However, we will see that the LO ChPT results work rather well, much better than naively anticipated for  such a small source-sink separation.

\subsection{The axial vector 3-pt functions}

\begin{figure}[t]
\begin{center}
$R_4(\vec{q},t,t')$ and $R^{\perp}_4(\vec{q},t,t')$\\[2ex]
\includegraphics[scale=0.85]{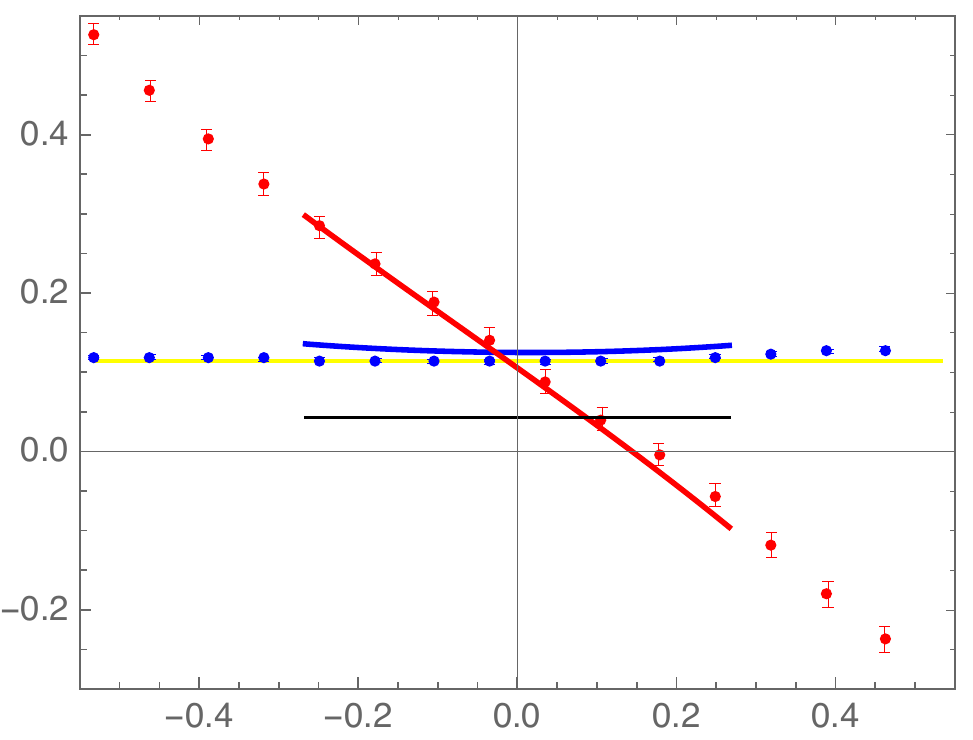}\\
$t'-t/2$ [fm]
\caption{\label{fig:A4} The correlation function ratios $R_4(\vec{q},t,t')$ (red) and $R^{\perp}_4(\vec{q},t,t')$ (blue) for $t= 1.07$ fm and a momentum transfer $Q^2=0.073\,{\rm GeV}^2$. The red and blue solid lines are the corresponding LO ChPT results. The black line shows the associated constant $\Pi_4(Q^2)$ the ratios converge to in the infinite $t$ limit. The yellow band indicates the ground state contribution obtained by fits to the data \cite{Bali:2018qus}.}
\end{center}
\end{figure}

Figure \ref{fig:A4} shows the data for $R_4(Q^2, t,t')$ (red data points) and $R^{\perp}_4(Q^2, t,t')$ (blue data points) as a function of the shifted operator insertion time $t'-t/2$ for fixed $t=1.07$ fm and momentum transfer $Q^2 =0.073$ GeV$^2$.\footnote{In Ref.\ \cite{Bali:2018qus} the data are displayed in figure 6, left panel.} 
The solid lines of the same color show the corresponding ChPT result. Apparently, LO ChPT describes the data very well. Recall that the ChPT results are not fits to the data, all input parameters are fixed as discussed in section \ref{SSect:Prelim}. 

The $R_4$ data do not exhibit a plateau and show, as a function of $t'$, an almost linear dependence with a large negative slope.
As explained in \cite{Bar:2018xyi}, this behaviour has two reasons: Firstly, the ground state SN matrix element is O($M_{\pi}/M_N$), while the $N\pi$ state matrix elements are O(1). Thus, the $N\pi$-state contamination is O($M_N/M_{\pi}$) enhanced compared to the SN matrix element. Secondly, one finds a relative sign between the coefficients $a^{\infty}_{4}(\vec{q})$ and $\tilde{a}^{\infty}_{4}(\vec{q})$ entering $Z^{\infty}_4$. This implies \cite{Bar:2018xyi}
\begin{equation}\label{Z4treeApprox}
Z^{\infty}_4(\vec{q},t,t') = -\frac{2M_NE_{\pi,\vec{q}}}{M_{\pi}^2} \exp\left(-\frac{E_{\pi,\vec{q}}\, t}{2}\right) \sinh\left(E_{\pi,\vec{q}}\left(t'-\frac{t}{2}\right)\right)\,,
\end{equation}
and it is essentially the $\sinh\left(E_{\pi,\vec{q}}\left(t'-\frac{t}{2}\right)\right)$ behaviour in this equation we observe in fig.\ \ref{fig:A4}. 
The prefactor in \pref{Z4treeApprox} is large because of the factor $M_N/M_{\pi}$. The large excited state  contamination in $R_4$ is one of the reasons why $R_4$ data are usually excluded from the calculation of the axial form factors.

According to \pref{DeltaZ4infty} the projection method removes the dominant $N\pi$ state contamination $Z^{\infty}_4$. The remaining contributions are O($M_{\pi}/M_N$) suppressed and therefore smaller. In addition, there is no relative sign between the coefficients $a^{\infty}_{k}(\vec{q})$ and $\tilde{a}^{\infty}_{k}(\vec{q})$, $k=1,2,3$, Therefore, the $R_4^{\perp}$ shows the familiar $\cosh$ behaviour. 
However, note that the $N\pi$ contamination is still rather large, even though the mild curvature in the $R^{\perp}_4$ data may suggest otherwise: The midpoint estimate $R^{\perp}_4(\vec{q}, t,t'=t/2)$ is a factor $\approx 2.9$ larger than the SN result, shown by the solid black line in fig.\ \ref{fig:A4}. This decreases to a factor $\approx1.8$ if the source-sink separation is increased by a factor 2 to $t\approx 2.1$ fm. The reason for this slow convergence is the smallness of the SN result. Large source-sink separations are necessary to exponentially suppress the O$(1)$ $N\pi$ contamination compared to the small SN contribution of O$(M_{\pi}/M_N)$.

As discussed before we expect a small impact of the projection method on the ratios $R_k$, $k=1,2,3$. Instead of considering these ratios we look directly at the impact on the effective axial form factors $\GA^{\rm eff}$ and $\GPt^{\rm eff}$, which are extracted from axial vector ratios with spatial components only. 

The ratio 
\begin{eqnarray}
r_{\rm PPD}(Q^2,t)& \equiv &  \frac{Q^2 +M_{\pi}^2}{4M_N^2} \frac{\GPt^{\rm est}(Q^2,t)}{\GA^{\rm est}(Q^2,t)}
\end{eqnarray} 
is introduced as an estimator for the validity of the PPD hypothesis. If the lattice estimates for the two form factors satisfy \pref{ppd1} this ratio assumes the constant value 1. 

\begin{figure}[t]
\begin{center}
$r_{\rm PPD}(Q^2,t)$ and $r^{\perp}_{\rm PPD}(Q^2,t)$\\[2ex]
\includegraphics[scale=0.85]{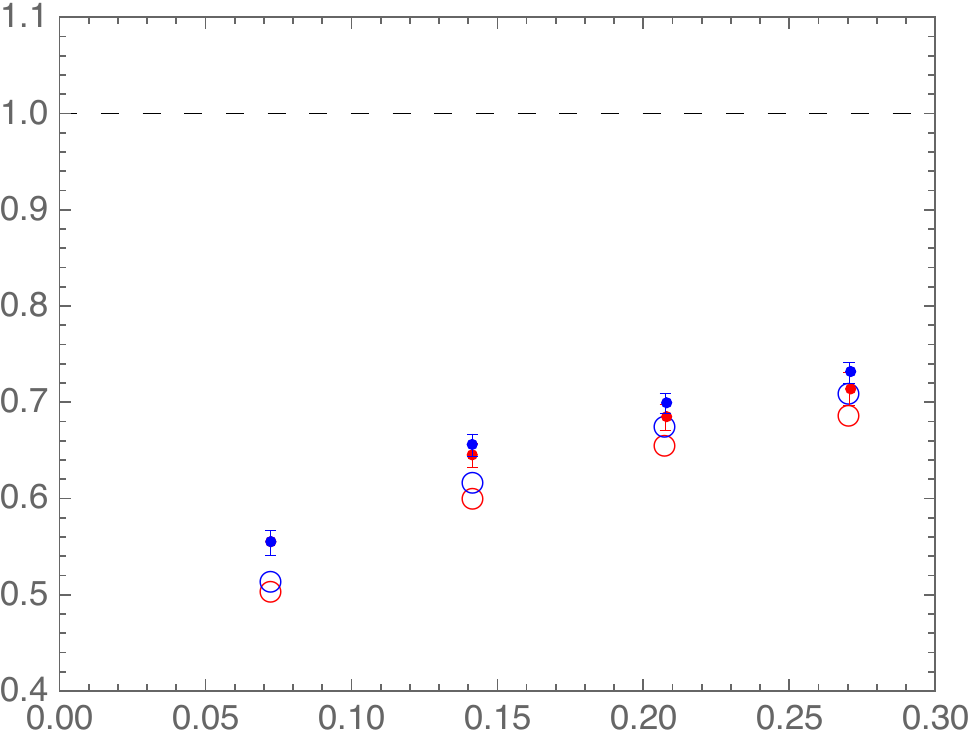}\\
$Q^2$ [GeV$^2$]
\caption{\label{fig:rPPD} RQCD data for $r_{\rm PPD}$ (red data points) and $r_{\rm PPD}^{\perp}$ (blue data points) for $t=1.07$ fm and the smallest four $Q^2$  accessible with $M_{\pi}=150$ MeV and $M_{\pi}L=3.5$. The open symbols (same color code) correspond to the ChPT results.}
\end{center}
\end{figure}

Figure \ref{fig:rPPD} shows the data for $r_{\rm PPD}$ (red data points) and $r^{ \perp}_{\rm PPD}$ (blue data points) for the smallest four momentum transfer accessible on Ensemble VIII \cite{Bali:2018qus}. Within the statistical errors there is no difference between the data for the projected and the standard axial vector currents. The open circles show the corresponding ChPT results (same color code) when the midpoint estimates for the axial and induced pseudoscalar form factors are used. Here too the symbols overlap and no significant difference is found. 

$r_{\rm PPD}$ is substantially smaller than 1, and the discrepancy increases the smaller $Q^2$ is. Thus, the PPD hypothesis seems strongly violated. As explained in Ref.\ \cite{Bar:2019gfx}, the dominant source is the $N\pi$ contamination in the induced pseudoscalar form factor. It results in a substantial underestimation of $\GPt$ that increases for small momentum transfer. 

Note that the PPD result \pref{ppd1} holds exactly in ChPT but does not need to hold in QCD. The small discrepancy between the lattice data and the ChPT results in fig.\ \ref{fig:rPPD} may be an indication for this. Still, it is remarkable how well the data is described by LO ChPT.

\subsection{The pseudoscalar 3-pt function}

\begin{figure}[p]
\begin{center}
$R_P(\vec{q},t,t')$ and $R^{\perp}_P(\vec{q},t,t')$\\[0.2ex]
\includegraphics[scale=0.85]{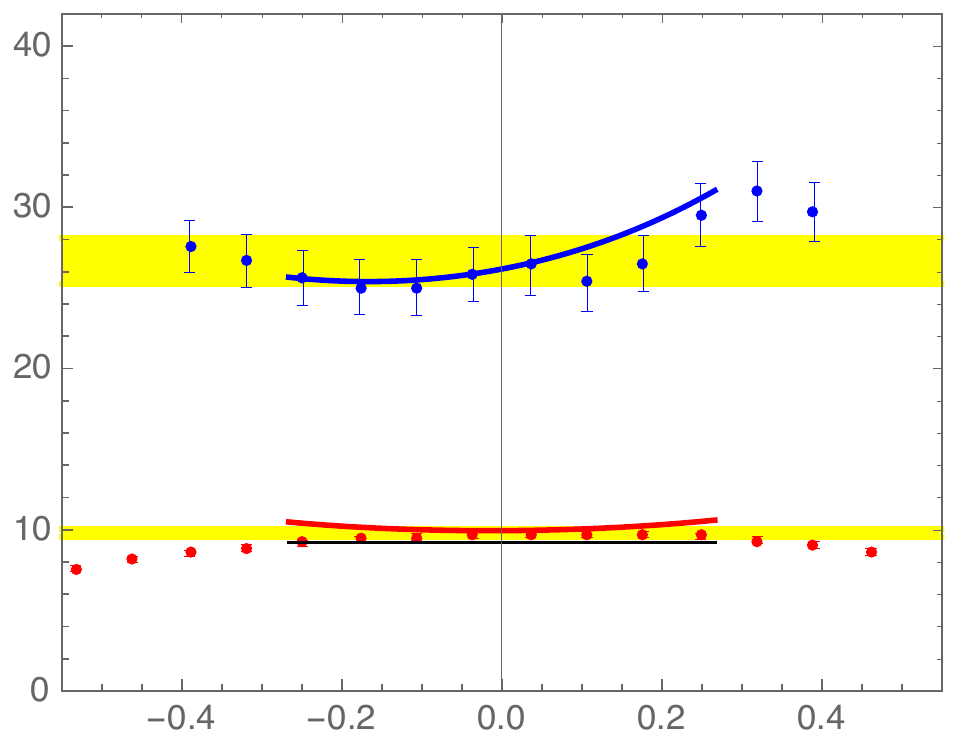}\\
$t'-t/2$ [fm]
\caption{\label{fig:RP} The correlation function ratios  $R_P(\vec{q},t,t')$ (red) and $R^{\perp}_P(\vec{q},t,t')$ (blue) for $t= 1.07$ fm and  $Q^2=0.073\,{\rm GeV}^2$. The red and blue solid lines are the corresponding LO ChPT results, the black line shows the associated constant $\Pi_P(Q^2)$ the ratios converge to in the infinite $t$ limit. The yellow bands indicate the ground state contributions obtained by fits to the data \cite{Bali:2018qus}.}
\vspace{1cm}
$Z_P(\vec{q},t,t')$ and $Z^{\perp}_P(\vec{q},t,t')$\\[0.2ex]
\includegraphics[scale=0.85]{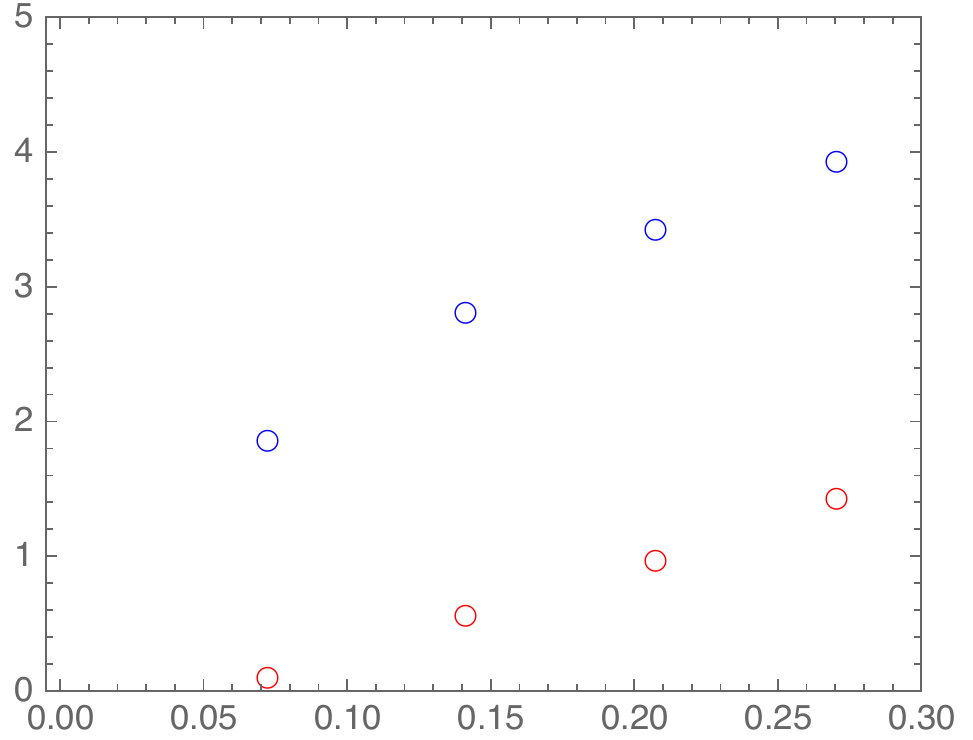}\\
$Q^2$ [GeV$^2$]
\caption{\label{fig:ZP} ChPT results for the $N\pi$ contaminations $Z_P(\vec{q},t,t')$ (red) and $Z^{\perp}_P(\vec{q},t,t')$ (blue) for $t= 1.07$ fm, $t'=t/2$ and 
the lowest four $Q^2$ accessible with $M_{\pi}=150$ MeV and $M_{\pi}L=3.5$. For the smallest momentum transfer $Z_{P}$ is accidentally close to zero (see main text). }
\end{center}
\end{figure}

Figure \ref{fig:RP} shows the data for $R_P(\vec{q}, t,t')$ (red data points) and $R^{\perp}_P(\vec{q}, t,t')$ (blue data points) as a function of  $t'-t/2$.\footnote{In Ref.\ \cite{Bali:2018qus} the data are displayed in figure 6, right panel.} 
The solid lines of the same color show the corresponding ChPT results.\footnote{For the $P^{a,\perp}$ result the quark mass $m_q$ and the renomalization factors $Z_A,Z_P$ are needed and taken from Ref.\ \cite{Bali:2014nma}, tables I - III.}  The yellow bands indicate the ground state contributions extracted in \cite{Bali:2018qus} by fits to the data. The black line shows the ChPT result for infinite source-sink separation, i.e.\ the ChPT result for the SN contribution. 

The LO ChPT results describe the data rather well.  Near the midpoint $t' \approx t/2$ the ChPT results are within the yellow bands. Besides this we observe two striking features in fig.\ \ref{fig:RP}: a) the $N\pi$ contamination in $R_P$ is tiny, the red and black lines are very close, and b), the data and the ChPT results for $R^{\perp}_P$ and $R_P$ differ vastly, roughly by a factor 3. Thus, the $N\pi$ contamination in $R^{\perp}_P$ is huge. 

Both features can be understood with the results presented in the last section.
According to \pref{DefZPinf} $Z^{\infty}_{P}$ is the sum of two contributions, the spatial one $\sum_{k=1}^3 \alpha_k Z_k^{\infty}$  and $Z^{\prime, \infty}_{4}$. While the spatial one is larger then zero, we find $Z^{\prime, \infty}_{4}<0$. The origin for the latter is visible in fig.\ \ref{fig:A4}: It stems from the time derivative $\partial_{t'}$ of $C_{3,A_4}(\vec{q},t,t')$, and $Z^{\prime}_{4}<0$ is nothing but the negative slope of $R_4(\vec{q},t,t')$ as a function of $t'$.

Due to the opposite sign the two contributions in $Z_{P}$ largely compensate. This compensation is not perfect, for small momentum transfers we find $Z_{P}<0$, while it turns positive for larger $Q^2$, see fig.\ 1 in Ref.\ \cite{Bar:2019gfx}. This implies a particular value for $Q^2$ where $Z_P$ vanishes. For $t=2$ fm and physical pion mass this value is approximately 0.065 GeV$^2$ \cite{Bar:2019gfx}, and it does not change much for the setup considered here. Accidentally, this value is close to the momentum transfer $Q^2=0.073\,{\rm GeV}^2$ underlying the data shown in fig.\ \ref{fig:RP}, thus explaining why the $N\pi$ contamination is so small in this figure. This coincidence is accidental, it stems from the particular setup with $M_{\pi}=150$ MeV and $M_{\pi}L=3.5$, which implies $Q^2=0.073\,{\rm GeV}^2$ for the smallest non-vanishing momentum $\vec{q}$ with $|\vec{q}|=2\pi/L$. 

As discussed in the previous section, the projection method subtracts $Z^{\prime, \infty}_{4}$ from the $N\pi$ contamination, leaving the large positive spatial contribution in $Z^{\perp,\infty}_P$. The partial removal by the projection method results in a large positive $N\pi$ contamination for the projected pseudoscalar density correlation function.

Figure \ref{fig:ZP} shows the $N\pi$ contaminations $Z_P(\vec{q},t,t')$ and $Z^{\perp}_P(\vec{q},t,t')$ at the midpoint $t'=t/2$ for the lowest four momentum transfers accessible.
The accidental $Z_P\approx 0 $ for the lowest $Q^2$ changes to non-vanishing positive values for the larger momentum transfers. Also the difference $\Delta Z_{P}$ increases for increasing $Q^2$.

\subsection{The generalized Goldberger-Treiman relation}

\begin{figure}[t]
\begin{center}
$r_{\rm PCAC}(Q^2,t)$ and $r^{\perp}_{\rm PCAC}(Q^2,t)$\\[0.2ex]
\includegraphics[scale=0.85]{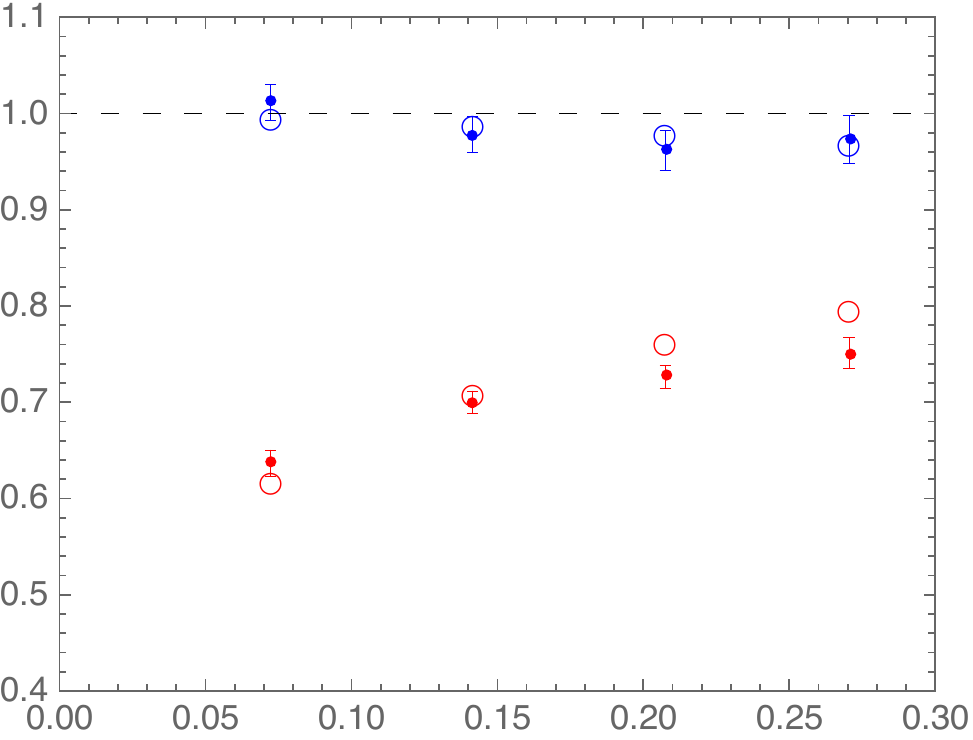}\\
$Q^2$ [GeV$^2$]
\caption{\label{fig:gGT} Data for the ratios $r_{\rm PCAC}(Q^2,t)$ (red) and $r^{\perp}_{\rm PCAC}(Q^2,t)$ (blue) for $t= 1.07$ fm and the lowest for momentum transfers $Q^2$ accessible with with $M_{\pi}=150$ MeV and $M_{\pi}L=3.5$. The corresponding LO ChPT are shown by the open circles (same color code).}
\end{center}
\end{figure}

Figure \ref{fig:gGT} shows the data for $r_{\rm PCAC}(Q^2,t)$ (red data points) and $r^{\perp}_{\rm PCAC}(Q^2,t)$ (blue data points) as a function of  $Q^2$ for $t=1.07$ fm. 
The open circles (same color code) are the corresponding LO ChPT results when the midpoint estimates for all three form factors are used.  Once again we find very good agreement between the lattice data and the ChPT results.

The lattice data for $r_{\rm PCAC}(Q^2,t)$ display the original PCAC puzzle. The ratio is smaller than 1 and the discrepancy increases the smaller the momentum transfer is. It was shown in Ref.\ \cite{Bar:2018xyi} that the $N\pi$ state contamination in $\GPt$ is the dominant source for this discrepancy. The induced pseudoscalar form factor is substantially underestimated due to the $N\pi$ contamination. Consequently, the first term on the right hand side of \pref{DefrPCAC} turns out to be too small.

In contrast, $r^{\perp}_{\rm PCAC}(Q^2,t)$ is close to 1.
With the findings of the last subsection the reason for this apparent improvement is easily identified. While the axial form factors $\GA^{\perp}, \GPt^{\perp}$ are essentially unchanged compared to $\GA,\GPt$, the pseudoscalar form factor $\GP^{\perp}$ receives a large $N\pi$ contamination leading to a significant overestimation of the pseudoscalar form factor.  This compensates the underestimation of $\GPt$ and results in $r^{\perp}_{\rm PCAC}(Q^2,t) \approx 1$. 

\section{Concluding remarks}

We can conclude that the projection method does not provide the desired solution for the excited-state contamination in nucleon axial form factors. Even though the generalized Goldberger-Treiman relation is satisfied
after the projection has been performed, the situation for the individual form factors is worse.
Instead of eliminating the large excited state contamination in the induced pseudoscalar form factor the projection method introduces an additional large one in the pseudoscalar form factor. These two essentially compensate each other in the measure $r_{\rm PCAC}$, and  $r_{\rm PCAC}\approx1$ falsely indicates the removal of all excited-state contaminations.

Notwithstanding the negative outcome for the projection method, the results of this paper strongly support the findings and conclusions of Refs.\ \cite{Bar:2018xyi,Bar:2019gfx}. ChPT is a useful tool to provide theoretical understanding for the $N\pi$ excited-state contamination in nucleon form factor calculations. The comparison between the LO ChPT results and lattice data works remarkably well, even at rather small euclidean time separations. This suggests that two-particle $N\pi$ states are responsible for the dominant excited-state contamination in lattice estimates for the form factors. Other excited states seem to have a small if not negligible impact.\footnote{The ChPT result for the three-particle $N\pi\pi$-state contamination in the nucleon 2-pt function is indeed found to be negligibly small \cite{Bar:2018wco}.} 

According to Ref.\ \cite{Bar:2018xyi} the large excited-state contamination in $\GPt$ stems from a particular low-momentum $N\pi$ state, where the axial vector current at $t'$ either directly creates a pion that propagates to the sink, or destroys a pion that was created at the source. The same state is responsible for the large $N\pi$ contamination in the ratio $R_4$. This has recently been exploited in \cite{Jang:2019vkm} to remove the excited-state contamination in $\GPt$ using $R_4$ data as input in the analysis of $\GPt$ data. Although this may turn out to be a viable method for $\GPt$ it is not expected to work for $\GA$. For this form factor the ChPT prediction for the $N\pi$ contamination is very different. Instead of stemming from one $N\pi$ state with a small pion momentum the $N\pi$ contamination is the cumulative contribution of many states that is not related to $R_4$ data. 

In a recent paper \cite{Bali:2019yiy} RQCD devised a new analysis strategy based on the theoretical insights obtained from the ChPT results. The method has been applied to the axial and pseudoscalar form factor calculations with encouraging results. The dominant $N\pi$-state contamination can be removed from the correlation functions and the SN ground state matrix elements can be extracted reliably.  The lattice result for the induced pseudoscalar coupling $g^{\ast}_P$ at the muon capture point is found to be in good agreement with the experimentally measured value. This warrants analogous ChPT calculations for other nucleon observables, for instance the nucleon electromagnetic form factors. 

\vspace{4ex}
\noindent {\bf Acknowledgments}
\vspace{2ex}

Discussions and correspondence with G.\ Bali, S.\ Collins, M.\ Gruber, R.\ Sommer, P.\ Wein and T.\ Wurm  are gratefully acknowledged.
I thank T.\ Wurm for sending me data published in Ref.\ \cite{Bali:2018qus}.
This work was supported by the German Research Foundation (DFG), Grant ID BA 3494/2-1.
\begin{appendix}

\section{The $N\pi$-contamination $Z_{P}^{\perp}$}\label{AppA}
In this appendix we derive the result for the $N\pi$ contamination $Z_{P}^{\perp}$ of the projected pseudoscalar density. 
The main task is to express the expressions in terms of the known results for the standard axial vector current.

The extra term in the projected pseudoscalar density \pref{DefPperp} involves the partial derivatives $\partial_{\mu}A_{\nu}^a(x)$ of the axial vector current. Computing the 3pt function \pref{C3ptP} with $P^{\perp}$ and comparing with \pref{DefZmuperp} we find 
\begin{eqnarray}\label{aux1DZP}
\Delta Z_{P}(\vec{q},t,t')&=&\sum_{\mu,\nu}\frac{\overline{p}_{\mu}\overline{p}_{\nu}}{\overline{p}^2}\, \frac{C^{N\pi}_{3,\partial_{\mu}A_{\nu}^a} (\vec{q},t,t')}{2M_N C^{N}_{3,A_{4}^a} (\vec{q},t,t')}\,.
\end{eqnarray}
Two comments are appropriate here. Firstly we have used the result $\sum_{\nu}\overline{p}_{\nu} C^{N\pi}_{3,\partial_{\mu}A_{\nu}^a} (\vec{q},t,t')=0$, a direct consequence of   $\sum_{\nu}\overline{p}_{\nu} C^{N\pi}_{3,A^a_{\nu}} (\vec{q},t,t')=0$ \cite{Bali:2018qus}. Secondly, the denominator in \pref{aux1DZP} stems from the replacement \cite{Bar:2019gfx}
\begin{equation}
2m C^{N}_{3,P^a} (\vec{q},t,t')= 2M_N C^{N}_{3,A_{4}^a} (\vec{q},t,t')\,.
\end{equation}
The 3-pt function of $\partial_{\mu}A_{\nu}^a$ in the numerator of \pref{aux1DZP} is related to the 3-pt function of the axial vector itself. Performing a partial integration we find
\begin{eqnarray}
C_{3,\partial_{\mu}A_{\nu}^a} (\vec{q},t,t')= 
\Bigg\{
\begin{array}{rcl}
\partial_{t'} C_{3,A^a_{\nu}}(\vec{q},t,t')\,,  &   & \mu=4,  \\[0.2ex]
-iq_k C_{3,A^a_{\nu}}(\vec{q},t,t')\,,  &   & \mu=k=1,2,3.  
\end{array}
\end{eqnarray}
Using this result in \pref{aux1DZP} the $N\pi$ contribution $ \Delta Z_{P}$ can be expressed in terms of the $N\pi$ contributions $Z_\mu$. The result simplifies if we take into account the NR expansion \pref{NRExpProjector} for $\overline{p}_{\mu}\overline{p}_{\nu}/\overline{p}^2$ up to O($1/M_N^2)$, leading to
\begin{eqnarray}\label{resDZP}
\Delta Z_{P}&=& Z_4^{\prime} - \sum^3_{k=1}\alpha_k Z_k^{\prime} + \frac{\vec{q}^{\,2}}{4M_N^2}\left(Z_4 - \sum^3_{k=1} \alpha_k Z_k\right)\,.
\end{eqnarray}
The $\alpha_k$ are the short hand notation for the combination
\begin{equation}\label{Defalphak}
\alpha_k(\vec{q}) =  -i  \frac{C_{3,A^a_k}^N(\vec{q},t,t')}{C_{3,A^a_4}^N(\vec{q},t,t')} \frac{q_k}{2M_N}\,,
\end{equation}
but performing the NR expansion we obtain the simple results \cite{Bar:2019gfx}
\begin{eqnarray}\label{resalphak}
\alpha_{k} & = &-\frac{q_k^2}{M_{\pi}^2}, \quad k=1,2\,, \qquad \alpha_{3} \, =\,\frac{\Epiq^2 -q_3^2}{M_{\pi}^2}\,.
\end{eqnarray}
The primed contributions $Z_{\nu}^{\prime}$ stem from the 3-pt function with the time derivative $\partial_{t'}$,
\begin{eqnarray}
Z_{\nu}^{\prime}(\vec{q},t,t')& =&  \frac{\partial_{t'} C^{N\pi}_{3,A^a_\nu}(\vec{q},t,t')}{2M_N C_{3,A^a_\nu}^N(\vec{q},t,t')}\,.
\end{eqnarray}
These have the same form as the original $Z_{\nu}(\vec{q},t,t')$, but with primed coefficients:
\begin{eqnarray}
Z_{\nu}^{\prime}(\vec{q},t,t') & = &  \phantom{+ \sum_{\vec{k}} }a^{\prime}_{\nu}(\vec{q}) e^{-\Delta E(0,\vec{q}) (t-t')}+ \tilde{a}^{\prime}_{\nu}(\vec{q}) e^{-\Delta E(\vec{q},-\vec{q})t'} \nn \\
& & + \sum_{\vec{k}} b^{\prime}_{\nu}(\vec{q},\vec{k}) e^{-\Delta E(0,\vec{k}) (t-t')}+\sum_{\vec{k}}\tilde{b}^{\prime}_{\nu}(\vec{q},\vec{k}) e^{-\Delta E(\vec{q},\vec{k}) t'} \nn\\
& & +  \sum_{\vec{k}} c_{\nu}^{\prime}(\vec{q},\vec{k}) e^{-\Delta E(0,\vec{k}) (t-t')}e^{-\Delta E(\vec{q},\vec{k}) t'}\,.
\label{DefC3Npcontrprime}
\end{eqnarray}
The primed coefficients involve additional factors stemming from the time derivative $\partial_{t'}$ of the exponentials in $C^{N\pi}_{3,A^a_\nu}(\vec{q},t,t')$:
\begin{eqnarray}
a^{\prime}_{\nu}(\vec{q}) &=&  \frac{E_{\pi,\vec{q}}}{2M_N} \,\,a_{\nu}(\vec{q}),\label{apP0}\\
\tilde{a}^{\prime}_{\nu}(\vec{q}) &=& -\frac{E_{\pi,\vec{q}}}{2M_N} \,\,\tilde{a}_{\nu}(\vec{q}),\label{atpP0}\\
b^{\prime}_{\nu}(\vec{q},\vec{k}) & =& \frac{E_{\pi,\vec{k}} +E_{N,\vec{k}} - E_{N,\vec{q}}}{2M_N} \,\,b_{\nu}(\vec{q},\vec{k}),\\
\tilde{b}^{\prime}_{\nu}(\vec{q},\vec{k}) &=& -\frac{E_{\pi,\vec{k}} -(E_{N,\vec{k}+\vec{q}} - E_{N,\vec{q}}) + (E_{N,\vec{q}} - M_N)}{2M_N} \,\, \tilde{b}_{\nu}(\vec{q},\vec{k}),\\
c_{\nu}^{\prime}(\vec{q},\vec{k}) & =& - \frac{E_{N,\vec{k}+\vec{q}} - E_{N,\vec{k}}}{2M_N}\,\,c_{\nu}(\vec{q},\vec{k}).\label{cpP0}
\end{eqnarray}
Note that the primed coefficients are $1/M_N$ suppressed relative to their unprimed counterparts. Thus, $Z^{\prime}_{\nu}$ contributes at one order higher in the NR expansion compared to $Z_{\nu}$. Therefore, to LO in the NR expansion eq.\ \pref{resDZP} reduces to the results \pref{DeltaZPinfty}, \pref{ZPinftyperp} presented in section \ref{ssect:Npicontr}.

\end{appendix}


\end{document}